\documentclass[floatfix,amsmath,amssymb,aps,prl,twocolumn]{revtex4}
\newcommand{\SPvec}[1]{{\textbf{\textsl{#1}}}}

\newcommand{\phiBI}{{\phi_{\beta}}}

\begin{document}
%
%	\draft              % \draft command makes pacs numbers print
%
	\title{Comment on ``Comment on {``Nonperturbative calculation of Born-Infeld effects on 
		the Schr\"odinger spectrum of the hydrogen atom''}'' by M. N. Smolyakov} 
\author{Holly K. Carley}
\email{HCarley@citytech.cuny.edu}
\affiliation{New York City College of Technology, CUNY, Department of Mathematics, 
                300 Jay Street, Brooklyn, NY 11201}
\author{Michael K.-H. Kiessling}
\email{miki@math.rutgers.edu}
\affiliation{Rutgers, The State University of New Jersey, 
	Department of Mathematics, 
	110 Frelinghuysen Rd., 
	Piscataway, NJ 08854}

\begin{abstract}
\noindent
{This reply to Dr. Smolyakov's comment \cite{COMMENT} 
on our PRL paper \cite{CarKiePRL} was solicited by PRL on Nov 17, 2021 and
submitted to PRL on Nov. 30, 2021. Regretfully, the editors of PRL decided not to publish Dr. Smolyakov's comment;
hence, not our reply to it. However, since Dr. Smolyakov made his comment publicly available at the arXiv.org in 
June 2022 (of which we learned only on Dec. 26, 2024), we have decided to make our reply to his comment
publicly available as well.}
\end{abstract}

\maketitle 

 Our paper \cite{CarKiePRL} has prompted a belated comment by Dr. Smolyakov \cite{COMMENT}, in which he writes:
 ``As will be demonstrated below, the solution for the electrostatic potential used in \cite{CarKiePRL} is not correct.'' 
 Since we in \cite{CarKiePRL} had stated that we were working with an {approximation} 
to the exact solution for the electrostatic potential in question,
we do not quite see the point of demonstrating that our approximation is not exact.
 Moreover, in the meantime systematic corrections to our approximation have been computed (see below). 
 It is interesting, though, to see that our approximation gives different results for two line
integrals that ordinarily coincide if no approximation is made. 
 Based on the nature of our approximation we had taken it for granted that this identity is preserved
under the approximation, and so, in \cite{CarKiePRL}, while talking about one line integral, we
tacitly worked with the other, computed earlier in \cite{mkB}.
 We thank Dr. Smolyakov for noticing that the two integrals give different results.

 However, the fact that these two line integrals yield different results under our approximation does not
invalidate our approximation method. 
 In \cite{KieJMP} and \cite{CarKieMPAG} we addressed the slightly simpler problem with regularized point 
charge densities and
showed that their standard Coulomb field $\SPvec{D}_{\mathrm{C}}(\SPvec{s})$ is the leading-order contribution to a 
convergent expansion of the electrostatic displacement field $\SPvec{D}(\SPvec{s})$ in powers of $\beta$; more
precisely,
$\SPvec{D}(\SPvec{s}) = \SPvec{D}_{\mathrm{C}}(\SPvec{s}) + \nabla\times \SPvec{G}(\SPvec{s})$ with 
 $\nabla\times \SPvec{G}(\SPvec{s}) = \sum_{k=1}^\infty \beta^{4k} \SPvec{D}^{(k)}(\SPvec{s})$, and we give
explicit integral representations of the $\beta$-independent solenoidal fields $\SPvec{D}^{(k)}(\SPvec{s})$. 
 We take this result as a vindication of our approximation method used in \cite{CarKiePRL},
where we had written:
``Assuming $\SPvec{D} = \SPvec{D}_{\mathrm{C}}$ in leading order in $\beta$, on this line 
we can approximately set $\phi (\SPvec{s}_e) = \phiBI (\SPvec{s}_e)$.''
 The line runs through the point charges, and $\phi (\SPvec{s}_e)$ is our approximation to 
the actual $\phiBI (\SPvec{s}_e)$, computed with $\SPvec{D}_{\mathrm{C}}$ in place of $\SPvec{D}$.

 In addition to demonstrating that corrections to the Coulomb field  $\SPvec{D}_{\mathrm{C}}$
of regularized point charge densities can be systematically computed via a series expansion, 
in \cite{mkCMPa} and \cite{mkCMPb} we showed that the sought-after electrostatic potential 
${\phiBI}(\SPvec{s}) \equiv \phiBI(\SPvec{s}{}|\SPvec{s}_p,\SPvec{s}_e)$ for two true point charges
that solves the nonlinear partial differential equation
\begin{equation}
-\nabla \cdot
   \frac{\nabla{\phiBI}(\SPvec{s})}
	{\sqrt{1-\beta^4|\nabla{\phiBI}(\SPvec{s})|^2 }}
=
4\pi \left(\delta_{{\scriptstyle\SPvec{s}}_p}(\SPvec{s})
-\delta_{{\scriptstyle\SPvec{s}}_e}\!(\SPvec{s})\right)
\label{eq:PHIstatic}
\end{equation}
 is the minimizer to the action functional 
\begin{equation}
 \mathcal{A}[\phi] =\!\! \int_{\mathbb{R}^3}\!\! \left(1- \sqrt{1-\beta^4|\nabla{\phi}|^2 } 
- 4\pi \beta^4\! \left(\delta_{{\scriptstyle\SPvec{s}}_p}
-\delta_{{\scriptstyle\SPvec{s}}_e}\right)\phi\right)\!d^3s
\label{eq:PHIfunc}
\end{equation}
under the asymptotic condition ${\phiBI}({\SPvec{s}})\to 0$ for $|{\SPvec{s}}|\to\infty$.
 The minimizer exists, is unique, and real analytic away from the location
of the (unit) point charges.
 This variational principle paves the ground for computing arbitrarily accurate numerical approximations
to $\phiBI(\SPvec{s}{}|\SPvec{s}_p,\SPvec{s}_e)$ with the help of powerful functional minimization algorithms.
 The computational effort reduces somewhat if one notices that, after at most a rotation and translation, one may assume
that the two point charges are located on the $z$ axis, with the $x,y$ plane halving the line segment between
the two point charges. 
 With this arrangement the minimizer is rotation invariant about the $z$ axis and reflection anti-symmetric across
the $x,y$ plane. 
 Yet even after this reduction to a minimization over (say) the first quadrant of the $x,z$ plane,
a significant amount of computations is required to  
minimize the functional for a sufficiently large interval of $\beta$ values and for a representative
number of distances $|\SPvec{s}_p -\SPvec{s}_e|=:r$ between the point charges so that the 
$\beta$-dependent Born--Infeld effects on the Schr\"odinger spectrum of hydrogen can be assessed with
the desired precision.

 We therefore agree with Dr. Smolyakov that it is an important open problem to carry out the accurate 
computation of the electrostatic potential $\phiBI(\SPvec{s})$ 
of two oppositely charged point particles in an otherwise empty electrostatic Born--Infeld vacuum,
with vanishing ``boundary'' conditions at spatial infinity $|\SPvec{s}|\to\infty$; cf. our 
endnote [20] in \cite{CarKiePRL}.
 Based on what we reported here we do not expect, though, that the conclusions of our paper \cite{CarKiePRL} 
would have to be revised, only to be refined. 
 This expectation is supported also by our investigation of the Bopp--Land\'{e}--Thomas--Podolsky (BLTP)
effects on the Schr\"odinger spectrum of hydrogen \cite{CKP}.
 In BLTP electrodynamics the electrostatic potential of two point charges is exactly 
known in terms of elementary functions,
which immensely simplifies the computation of the Schr\"odinger eigenvalues.
 The conclusions arrived at in \cite{CKP} point in the same direction for the BLTP theory as those in \cite{CarKiePRL}
for the Born--Infeld theory.

 Lastly, we take the opportunity to point out that Franklin and Garon \cite{FG} reproduced table 1 of \cite{CarKiePRL}
except for the two numbers in column four, rows two and three. 
 In \cite{CKP} we reported that this numerical error was caused by a factor 2 mistake in our eigenvalue program, 
which showed only when the angular momentum quantum number $\ell>0$. 
 That correction did not change the conclusions of \cite{CarKiePRL}.

\end{document}